\documentclass[lettersize,journal]{IEEEtran}
\usepackage{amsmath,amsfonts}
\usepackage{algorithmic}
\usepackage{algorithm}
\usepackage{array}
\usepackage{textcomp}
\usepackage{stfloats}
\usepackage{url}
\usepackage{color}
\usepackage[none]{hyphenat}
\usepackage{verbatim}
\usepackage{graphicx}
\usepackage{float}  %设置图片浮动位置的宏包
\usepackage{subfigure}  %插入多图时用子图显示的宏包
\usepackage{booktabs} 
\usepackage{cite}
\usepackage{csquotes}
\begin{document}

\title{Security in Low-Altitude ISAC with Coupled Communication and Sensing Information Leakage}
\author{Meiding~Liu,~\IEEEmembership{Graduate Student Member,~IEEE},
		Yuyuan~Fang,~\IEEEmembership{Member,~IEEE},
        Zhengchun~Zhou,~\IEEEmembership{Member,~IEEE},
        Qiao~Shi,~\IEEEmembership{Member,~IEEE},
		and
        Pingzhi Fan,~\IEEEmembership{Life Fellow,~IEEE}

        \thanks{Meiding Liu, Yuyuan Fang, Zhengchun Zhou and Qiao Shi are with the School of Information Science and Technology, Southwest Jiaotong University, Chengdu 611756, China (e-mail:lmd@my.swjtu.edu.cn; fangyy@swjtu.edu.cn; zzc@swjtu.edu.cn; qiaoshi@swjtu.edu.cn).}
        \thanks{Pingzhi Fan is with the Key Laboratory of Information Coding and Transmission, Southwest Jiaotong University, Chengdu 611756, China (e-mail:pzfan@swjtu.edu.cn).}
}

\maketitle

\begin{abstract}
Secure information transmission in integrated sensing and communication (ISAC) systems deployed in low-altitude wireless networks (LAWN) is particularly challenging due to the open airspace and the broadcast nature of wireless signals reused for both communication and sensing, which give rise to more severe and complex security threats. This paper investigates the joint communication and sensing security of low-altitude ISAC systems where a communication eavesdropper (CEve) cooperates with and assists a sensing eavesdropper (SEve), leading to coupled communication and sensing information leakage. We define a recoverable signal power ratio (RSR) that quantifies the degree of coupling between communication eavesdropping and sensing information leakage. Based on this coupling, we formulate a joint eavesdropping mutual information (MI) that facilitates unified analysis and optimization for the overall security. Subsequently, we design the transmit beamformer that maximizes the weighted sum of legitimate communication and sensing MI under the joint information-leakage constraint. Simulations show that the proposed scheme converges rapidly and effectively exploits the coupling between communication eavesdropping capability and sensing information leakage.
\end{abstract}


\section{Introduction}
Integrated sensing and communication (ISAC) integrates sensing and communication functionalities through shared wireless resources and has been widely recognized as a key enabling technology for sixth-generation systems [1]. Nevertheless, the reuse of information-bearing signals for sensing purposes exposes ISAC systems to increased security risks, as confidential information may be leaked to malicious eavesdroppers (Eves) \cite{su}. This issue is particularly critical in low-altitude wireless networks (LAWN) due to the open airspace and the broadcast nature of wireless transmission \cite{lows}.

Information security in ISAC systems includes communication security and sensing security. The former focuses on preventing confidential information embedded in ISAC signals from being eavesdropped by communication Eves (CEves). This issue has been widely studied by utilizing physical layer security technology, where communication secrecy is typically quantified using information-theoretic metrics such as mutual information (MI) or secrecy rate (SR). To this end, researchers have enhanced secrecy performance through techniques such as secure beamforming, artificial noise-aided transmission and unmanned aerial vehicle (UAV)-assisted schemes\cite{su, collude, Wu2026},\cite{Wu2025}.

In addition to communication security, ISAC systems raise sensing security concerns \cite{WinWin, radarp}, as sensing Eves (SEves) can infer sensitive information about the targets, such as location and velocity, from received echo signals, resulting in privacy leakage. To address this, \cite{Zou} considered a SEve with perfect knowledge of the transmitted signals and designed the beamformers to maximize the authorized sensing MI while limiting the sensing eavesdropping MI. Moreover, \cite{UAV} studied a UAV-enabled secure ISAC system with a dual Eve in LAWN, and jointly designed the beamformers and UAV trajectory to maximize the legitimate communication SR, while ensuring sensing security by limiting signal power in the direction of the Eve. Besides, \cite{ren1} considered a cell-free ISAC system with SEve and CEve, where the SEve lacks knowledge of the transmitted signals and relies on non-coherent energy detection, and the beamformers are designed to maximize the legitimate sensing detection probability, while constraining the eavesdropping probability at the SEve and satisfying communication signal-to-interference-plus-noise ratio (SINR) constraints. 

Existing ISAC studies involving sensing security mainly focus on individual Eves and assume either perfect knowledge of the transmitted signals at the SEve or no access to them at all. In practice, especially in LAWN, the open and broadcast wireless environment, together with channel noise, interference, and potential cooperation among Eves, results in the SEve being able to only partially access the transmitted signals, which may still indirectly leak sensing information. Such situations may arise in practical low-altitude applications, such as UAV delivery, urban air mobility, and infrastructure inspection, where malicious nodes may cooperate to share intercepted signals \cite{sun}. These conditions invalidate these extreme assumptions, making them inadequate for quantifying coupled communication and sensing information leakage and complicating secure performance evaluation. Therefore, quantifying and mitigating this coupled leakage is essential to assess overall security risks and design secure transmission schemes in low-altitude ISAC systems with dual Eves.

In this paper, motivated by the above discussion, we consider a secure low-altitude ISAC system with cooperative CEve and SEve, where the CEve eavesdrops on the confidential transmission and forwards recovered signals to assist the SEve, leading to coupled communication and sensing information leakage. We define a recoverable signal power ratio (RSR) to quantify the coupling between communication eavesdropping and sensing information leakage. Based on this coupling, we construct a joint eavesdropping MI metric to evaluate the overall eavesdropping capability. Then, we formulate a transmit beamforming optimization problem that maximizes a weighted sum of the legitimate communication and sensing MI, subject to a unified information-leakage constraint limiting the eavesdropping capability. Simulation results demonstrate the impact of the RSR in coupling partial communication eavesdropping with sensing information leakage, and show that the proposed scheme enables an effective tradeoff between joint information leakage and legitimate performance.
%------------------------------------------------------------------------------------------
%------------------------------------------------------------------------------------------
\section{System Model and Performance Metrics}\label{2jie}    
As shown in Fig. 1, we consider a secure low-altitude ISAC system where a base station (BS), equipped with half-wavelength spaced uniform linear arrays of $M$ transmitting antennas and $N_{\mathrm r}$ receiving antennas, transmits dual-functional signals to serve a single-antenna communication user, and to sense an aerial target of interest simultaneously in the presence of cooperative CEve and SEve. Specifically, the CEve eavesdrops the confidential communication information and forwards its recovered signal estimate to assist the SEve, which thus obtains side information regarding the transmitted signal and exploits it to illegally infer the target-related sensing parameters. Moreover, the CSI of all legitimate and eavesdropping links is assumed to be perfectly available at the BS.
\begin{figure}[htbp]
	\centering
	\includegraphics[width=3in]{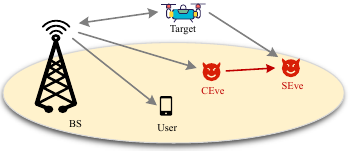}
	\caption{The secure ISAC system model with cooperative CEve and SEve.}
	\label{fig_1}
\end{figure}
\subsection{Legitimate Communication and Sensing}
The received signal at the user is expressed as 
\begin{align}
\label{1}
y_{\rm{U}} = {\bf{h}}_{\rm{U}}^{\rm{H}}{\bf{x}} + {n_{\rm{U}}},
\end{align}
where ${{\bf{h}}_{\rm{U}}} \in {\mathbb{C}^{M \times 1}}$ is the channel vector from the BS to the user, which is assumed to be perfectly known at the BS. Moreover, ${\bf{x}} = {\bf{w}}s \in {\mathbb{C}^{M \times 1}}$ is the transmitted signal vector, where $s$ is the information symbol with $\mathbb{E}\left[ {s{s^{\rm{H}}}} \right]=1$, and ${\bf{w}}$ is the transmit beamformer. Additionally, $n_{\rm{U}} \sim {\mathcal{CN}}(0,{\sigma_{\rm{U}} ^2})$ is the additive white Gaussian noise (AWGN) at the user. Then, the achievable rate of the user can be expressed as \cite{MuIn}
\begin{align}
\label{2}
{I_{\rm{U}}} = I\left( {{\bf{x}};{y_{\rm{U}}}\left| {{{\bf{h}}_{\rm{U}}}} \right.} \right) = {{\log }_2}\left( {1 + \sigma _{\rm{U}}^{ - 2}|{\bf{h}}_{\rm{U}}^{\rm{H}}{\bf{w}}{|^2}} \right).
\end{align}

For radar sensing, the main task is to identify the target-related parameters from the received echoes, which result from the reflection of the transmitted signal through the sensing channel. Accordingly, the sensing received signal can be expressed as
\begin{align}
{{\bf{y}}_{\rm{r}}} = {{\bf{H}}_{\rm{r}}}{\bf{x}} + {{\bf{n}}_{\rm{r}}},
\end{align}
where ${\bf H}_{\rm r}$ denotes the sensing channel matrix. We assume that 
$\mathrm{vec}({\bf H}_{\rm r})$ follows a zero-mean circularly symmetric complex Gaussian (CSCG) distribution with covariance matrix 
${\bf R}_{h_r} \in \mathbb{C}^{(MN_r)\times(MN_r)}$. 
Moreover, ${\bf{n}}_{\rm{r}} \sim {\mathcal{CN}}(0,{\sigma_{\rm{r}} ^2}{\bf{I}}_{N_r})$ is the AWGN at the BS.

To quantify the sensing performance at the receiver, we adopt the MI between the received radar signal and the sensing channel. This metric characterizes the amount of information that the received signal carries about the sensing channel and thus reflects the sensing capability. Accordingly, a higher sensing MI generally corresponds to improved performance in target parameter estimation, classification, and identification. The sensing MI is given by
\begin{align}
\begin{array}{l}
	\!\!\!\!\!\!{I_{\rm{r}}}\! =\! I\left( {{{\bf{H}}_{\rm{r}}};{{\bf{y}}_{\rm{r}}}\left| {\bf{x}} \right.} \right)
	\!= \! \log \det \!\left( {{\bf{I}}\! +\! \sigma _{\rm{r}}^{ - 2}{{\bf{R}}_{{h_{\rm{r}}}}}\!\left( {{\bf{w}}{{\bf{w}}^{\rm{H}}} \otimes {{\bf{I}}_{N{\rm{r}}}}} \right)} \right).
\end{array}
\end{align}
\subsection{Cooperative Communication and Sensing Eavesdropping}
In this subsection, we characterize the coupling between communication eavesdropping and sensing information leakage by deriving a closed-form expression for the sensing eavesdropping MI and further defining a joint eavesdropping MI based on the RSR.

We begin with the communication eavesdropping at the CEve. The received signal is given by
\begin{align}
{y_{{\rm{EC}}}} = {\bf{h}}_{{\rm{EC}}}^{\rm{H}}{\bf{x}} + {n_{\rm{EC}}},
\end{align}
where ${\bf h}_{\rm EC}\in\mathbb C^{M\times 1}$ is the communication eavesdropping channel vector and $n_{\rm EC}\sim\mathcal{CN}(0,\sigma_{\rm EC}^2)$ denotes the AWGN. Accordingly, the MI at CEve is
\begin{align}
{I_{{\rm{EC}}}} = I\left( {{\bf{x}};{y_{{\rm{EC}}}}\left| {{\bf{h}}_{{\rm{EC}}}^{}} \right.} \right) = {\log _2}( {1 + \sigma _{{\rm{EC}}}^{ - 2} {{{\left| {{\bf{h}}_{{\rm{EC}}}^{\rm{H}}{\bf{w}}} \right|}^2}} } ).
\end{align}
In the considered cooperative eavesdropping scenario, the CEve not only intercepts the confidential communication but also forwards its recovered symbol estimates to assist the SEve in sensing inference. As a result, the sensing information leakage at the SEve is influenced by the symbol recovery accuracy achieved at the CEve. Under the Minimum Mean Square Error (MMSE) estimation criterion, the transmitted signal can be decomposed as
\begin{align}
{\bf{x}} = \underbrace {{\bf{w}}\hat s}_{{{\bf{x}}_{{\rm{known}}}}} + \underbrace {{\bf{w}}\left( {s - \hat s} \right)}_{{{\bf{x}}_{{\rm{unk}}}}},
\end{align}
where $\hat s = \mathbb{E}\left[ {s|{y_{{\rm{EC}}}}} \right]$ denotes the MMSE estimate of the transmitted symbol at the CEve, $\mathbb{E}\left[  \cdot  \right]$ is the expectation operation, and $\tilde s = s - \hat s$ represents the residual uncertainty due to imperfect symbol recovery. Accordingly, ${\bf x}_{\rm known}$ represents the recoverable part available at the SEve through cooperation, while ${\bf x}_{\rm unk}$ is unavailable, being treated as the AWGN. By the orthogonality principle, the two components are statistically orthogonal, i.e., $\mathbb{E}[{{\bf x}_{\rm known}}{\bf x}_{\rm unk}^{\rm{H}}] = {\bf{0}}$ \cite{esti}.

Based on this, the received signal at the SEve can be expressed as 
\begin{align}
	{y_{{\rm{ES}}}} = {\bf{h}}_{{\rm{ES}}}^{\rm{H}}{{\bf{x}}_{{\rm{known}}}} + {\bf{h}}_{{\rm{ES}}}^{\rm{H}}{{\bf{x}}_{{\rm{unk}}}} + {n_{\rm{ES}}},
\end{align}
where ${{\bf{h}}_{{\rm{ES}}}} \in {\mathbb{C}^{M \times 1}}$ is the sensing eavesdropping channel, $n_{\rm{ES}} \sim {\mathcal{CN}}(0,{\sigma_{\rm{ES}} ^2})$ is the AWGN at the SEve. Based on the side information obtained from the recoverable signal, the conditional sensing eavesdropping MI is given by
\begin{align}
	\label{SMI}
\begin{array}{l}
{I_{{\rm{ES}}}} = I\left( {{{\bf{h}}_{{\rm{ES}}}};{y_{{\rm{ES}}}}\left| {{{\bf{x}}_{{\rm{known}}}}} \right.} \right)\\
~~~~~= h\left( {{y_{{\rm{ES}}}}\left| {{{\bf{x}}_{{\rm{known}}}}} \right.} \right) - h\left( {{y_{{\rm{ES}}}}\left| {{{\bf{h}}_{{\rm{ES}}}},{{\bf{x}}_{{\rm{known}}}}} \right.} \right),
\end{array}
\end{align}
which quantifies the information that the received signal carries about the sensing eavesdropping channel. To obtain a closed-form expression for $I_{\rm ES}$, we first derive the conditional variances corresponding to the two entropy terms in (\ref{SMI}), based on which the conditional entropies are obtained. Since ${\bf x}_{\rm known}={\bf w}\hat s$, conditioning on ${\bf x}_{\rm known}$ is equivalent to conditioning on $\hat s$. Accordingly, the conditional variance of $y_{\rm ES}$ for the first entropy term $h(y_{\rm ES}|{\bf x}_{\rm known})$ is given by
\begin{align}
	\begin{array}{l}
		\!\!\!\!\!\! \mathbb{E}\left[ {{{\left| {{y_{{\rm{ES}}}}} \right|}^2}\left| {\hat s} \right.} \right] 
		\mathop = \limits^{{\rm{(a)}}}
		\mathbb{E}\left[ {{{\left| {{\bf{h}}_{{\rm{ES}}}^{\rm{H}}{\bf{w}}\hat s} \right|}^2}} \right] 
		+ \mathbb{E}\left[ {{{\left| {{\bf{h}}_{{\rm{ES}}}^{\rm{H}}{\bf{w}}\tilde s} \right|}^2}} \right] 
		+ \sigma _{{\rm{ES}}}^2\\
		\!\!\!\!\!\!
		= {\rm{Var}}\left( {\hat s} \right){{\bf{w}}^{\rm{H}}}{{\bf{R}}_{{\rm{ES}}}}{\bf{w}} 
		+ {\rm{Var}}\left( {\tilde s} \right){{\bf{w}}^{\rm{H}}}{{\bf{R}}_{{\rm{ES}}}}{\bf{w}} 
		+ \sigma _{{\rm{ES}}}^2,
	\end{array}
\end{align}
where (a) follows from the independence of ${\bf h}_{\rm ES}$ from both $\hat s$ and $\tilde s$, with the MMSE orthogonality condition $E\left[ {\hat s{{\tilde s}^ * }} \right] = 0$, ${\bf{R}}_{{\rm{ES}}}$ denotes the sensing channel covariance matrix. Then, the first conditional entropy can be expressed as
\begin{align}
	\label{en1}
    \begin{array}{*{20}{l}}
	\!\!\!\!{h\left( {{y_{{\rm{ES}}}}\left| {{{\bf{x}}_{{\rm{known}}}}} \right.} \right) = {{\log }_2}\left( {\pi {\rm{e}}\mathbb{E}\left[ {{{\left| {{y_{{\rm{ES}}}}} \right|}^2}\left| {\hat s} \right.} \right]} \right)}\\
	\!\!\!\!{ = {{\log }_2}\left( {\pi {\rm{e}}\left( {\left( {{\rm{Var}}\left( {\hat s} \right) + {\rm{Var}}\left( {\tilde s} \right)} \right){{\bf{w}}^{\rm{H}}}{{\bf{R}}_{{\rm{ES}}}}{\bf{w}} + \sigma _{{\rm{ES}}}^2} \right)} \right).}
    \end{array}
\end{align}
For the second entropy term $h\left( {{y_{{\rm{ES}}}}\left| {{{\bf{h}}_{{\rm{ES}}}},{{\bf{x}}_{{\rm{known}}}}} \right.} \right)$, the remaining uncertainty arises from ${{\bf{h}}_{{\rm{ES}}}^{\rm{H}}{\bf{w}}\tilde s}$ and $n_{\rm ES}$. Then, the corresponding conditional variance is given by
\begin{align}
\!\! \mathbb{E}\left[ {{{\left| {{\bf{h}}_{{\rm{ES}}}^{\rm{H}}{\bf{w}}\tilde s + {n_{{\rm{ES}}}}} \right|}^2}\left| {{{\bf{h}}_{{\rm{ES}}}}} \right.} \right]\mathop  = \limits^{{\rm{(b)}}} {\rm{Var}}\left( {\tilde s} \right){{\bf{w}}^{\rm{H}}}{{\bf{R}}_{{\rm{ES}}}}{\bf{w}} + \sigma _{{\rm{ES}}}^2,
\end{align}
where (b) follows from the independence between $\tilde s$ and $n_{\rm ES}$. Accordingly, the second conditional entropy is 
\begin{align}
	\label{en2}
\begin{array}{l}
	h\left( {{y_{{\rm{ES}}}}\left| {{{\bf{h}}_{{\rm{ES}}}},{{\bf{x}}_{{\rm{known}}}}} \right.} \right) \!=\! {\log _2}\left( {\pi {\rm{e}}\mathbb{E}\left[ {{{\left| {{\bf{h}}_{{\rm{ES}}}^{\rm{H}}{\bf{w}}\tilde s +\! {n_{{\rm{ES}}}}} \right|}^2}\left| {{{\bf{h}}_{{\rm{ES}}}}} \right.} \right]} \right)\\
	= {\log _2}\left( {\pi {\rm{e}}\left( {{\rm{Var}}\left( {\tilde s} \right){{\bf{w}}^{\rm{H}}}{{\bf{R}}_{{\rm{ES}}}}{\bf{w}} + \sigma _{{\rm{ES}}}^2} \right)} \right).
\end{array}
\end{align}
By substituting (\ref{en1}) and (\ref{en2}) into (\ref{SMI}), the sensing eavesdropping MI at the SEve can be expressed as 
\begin{align}
	\label{IES2}
{I_{{\rm{ES}}}} = \log_2 \left( {1 + \frac{{{\rm{Var}}\left( {\hat s} \right){{\bf{w}}^{\rm{H}}}{{\bf{R}}_{{\rm{ES}}}}{\bf{w}}}}{{{\rm{Var}}\left( {\tilde s} \right){{\bf{w}}^{\rm{H}}}{{\bf{R}}_{{\rm{ES}}}}{\bf{w}} + \sigma _{{\rm{ES}}}^2}}} \right).
\end{align}
Since $I_{{\rm{ES}}}$ involves the recoverable signal power characterized by ${\rm{Var}}\left( {\hat s} \right)$, we first derive the variance of the MMSE estimate $\hat s$. Specifically, $\hat{s}$ is given by
		\begin{align}
			\hat s = \frac{({\bf h}_{\rm EC}^{H}{\bf w})^{*}} {|{\bf h}_{\rm EC}^{H}{\bf w}|^2+\sigma_{\rm EC}^2} y_{\rm EC},
		\end{align}
		which yields
		\begin{align}
			{\rm Var}(\hat s)= \frac{|{\bf h}_{\rm EC}^{H}{\bf w}|^2} {|{\bf h}_{\rm EC}^{H}{\bf w}|^2+\sigma_{\rm EC}^2}.
		\end{align}
Based on the variance of the MMSE estimate, we define the RSR as the fraction of the transmitted signal power that can be recovered at the CEve, i.e.,
\begin{align}
	\label{alpha}
	\alpha  = \frac{{{\rm{Var}}\left( {\hat s} \right)}}{{{\rm{Var}}\left( s \right)}} = \frac{{{{\bf{w}}^{\rm{H}}}{\bf{Aw}}}}{{{{\bf{w}}^{\rm{H}}}{\bf{Aw}} + 1}},0 \le \alpha  \le 1,	
\end{align}
where ${\bf{A}} = {\bf{h}}_{{\rm{EC}}}^{}{\bf{h}}_{{\rm{EC}}}^{\rm{H}}/\sigma _{{\rm{EC}}}^2$, and $\alpha$ quantifies the symbol recovery accuracy at the CEve. A larger $\alpha$ indicates better recovery performance, and $\alpha$ is determined by the signal-to-noise ratio at the CEve.

Substituting (\ref{alpha}) into (\ref{IES2}), the sensing eavesdropping MI is rewritten as
\begin{align}
\label{ies}
\begin{array}{l}
	{I_{{\rm{ES}}}}= {\log _2}\left( {1 + \dfrac{{\alpha {{\bf{w}}^{\rm{H}}}{{\bf{R}}_{{\rm{ES}}}}{\bf{w}}}}{{(1 - \alpha ){{\bf{w}}^{\rm{H}}}{{\bf{R}}_{{\rm{ES}}}}{\bf{w}} + \sigma _{{\rm{ES}}}^2}}} \right),
\end{array}
\end{align}
which highlights the explicit role of the communication recovery accuracy in the sensing eavesdropping MI.

\textit{Lemma~1: }(Monotonicity): The sensing eavesdropping MI at the SEve is monotonically non-decreasing with respect to the RSR $\alpha  \in \left[ {0,1} \right]$.

\textit{Proof:} Taking the derivative of $I_{{\rm{ES}}}$ in (\ref{ies}) with respect to $\alpha$, we obtain
\begin{align}
\frac{{\partial {I_{{\rm{ES}}}}}}{{\partial \alpha }} = \frac{{{{\bf{w}}^{\rm{H}}}{{\bf{R}}_{{\rm{ES}}}}{\bf{w}}}}{{\ln 2\left( {(1 - \alpha ){{\bf{w}}^{\rm{H}}}{{\bf{R}}_{{\rm{ES}}}}{\bf{w}} + \sigma _{{\rm{ES}}}^2} \right)}}.
\end{align}
Since ${{\bf{w}}^{\rm{H}}}{{\bf{R}}_{{\rm{ES}}}}{\bf{w}} \ge 0$, 
$\sigma _{{\rm{ES}}}^2 > 0$, and $\alpha \in [0,1]$, 
the numerator is non-negative and the denominator is strictly positive. Thus, 
$\partial I_{{\rm{ES}}}/\partial \alpha \ge 0$, implying that $I_{{\rm{ES}}}$ is monotonically non-decreasing with respect to $\alpha$. This completes the proof.

\textit{Remark~1.} Lemma 1 establishes a direct coupling between the communication eavesdropping capability at the CEve and the sensing information leakage at the SEve in cooperative eavesdropping. In particular, suppressing the recoverable signal power at the CEve inherently constrains the sensing performance of the SEve.

\textit{Lemma~2: }(Boundary Cases): When the RSR $\alpha$ satisfies\\
(1) $\alpha=0$, then ${I_{\rm{ES}}} = 0$.\\
(2) $\alpha=1$, then the sensing eavesdropping MI reduces to
\begin{align}
	\label{bound2}
{I_{{\rm{ES}}}} = {\log _2}\left( {1 + \sigma _{{\rm{ES}}}^{ - 2}{{\bf{w}}^{\rm{H}}}{{\bf{R}}_{{\rm{ES}}}}{\bf{w}}} \right),
\end{align}
which corresponds to the case where the transmitted signals are perfectly known at the SEve.

\textit{Proof:}
For $\alpha=0$, substituting it into (\ref{ies}) gives $I_{\rm ES}=\log_2(1)=0$. For $\alpha=1$, substituting it into (\ref{ies}) yields (\ref{bound2}). This completes the proof.

Finally, to quantify the overall information leakage under the dual-eavesdropping scenario, we define the joint eavesdropping MI\footnote{The proposed $I_{\rm DE}$ is introduced as a unified information-leakage metric rather than the MI of the joint observations of the two Eves. Although the CEve and SEve observations are statistically coupled through the recovered side information, $I_{\rm EC}$ and $I_{\rm ES}$ quantify leakage regarding different information objects. Their sum is therefore adopted as a tractable metric for evaluating the overall eavesdropping capability and facilitating secure beamforming design.} as
\begin{align}
I_{{\rm{DE}}} = {I_{{\rm{EC}}}} + {I_{{\rm{ES}}}},
\end{align}
which characterizes the coupled communication and sensing information leakage induced by the RSR, and serves as a unified performance metric for secure ISAC design.

\textit{Extension to Multi-User Scenarios:} The proposed framework can be extended to multi-user scenarios by generalizing the transmitted signal as ${\bf x}=\sum_{k=1}^{K}{\bf w}_k s_k$, where ${\bf w}_k$ and $s_k$ denote the beamformer and information symbol for the $k$-th user, respectively. In this case, each user's communication MI should account for both the desired signal and inter-user interference. The MMSE-based decomposition and RSR can then be generalized for each user, thereby extending the proposed joint communication and sensing information-leakage framework to multi-user scenarios.

\section{Beamforming Design under Joint Communication and Sensing Eavesdropping}\label{3jie}
In this section, we first formulate the secure beamforming optimization problem to maximize the legitimate communication and sensing performance while suppressing the information leakage to the Eves. Then, an efficient algorithm is developed to solve it. Finally, the convergence behavior and computational complexity are analyzed.
\subsection{Problem Formulation and Algorithm Design} 
The secure beamforming design problem is formulated as 
\begin{align}
	\label{12}
	&{\mathop {\max }\limits_{\bf{W}} ~{I_{\rm{U}}} + \beta {I_{\rm{r}}}}\tag{22a}\\
	&~{\rm{s}}.{\rm{t}}.~~{I_{{\rm{DE}}}} \le {\Gamma _{\rm{E}}}, \tag{22b}\\
	{}&{~~~~~~~{\rm{Tr}}\left( {\bf{W}} \right) \le {P_{\max }}},\tag{22c}\\
	{}&{~~~~~~~{\bf{W}}\succeq 0,{\rm{rank}}({\bf{W}}) = 1},\tag{22d}
\end{align}
where ${\bf{W}} = {\bf{w}}{{\bf{w}}^{\rm{H}}}$, $\beta$ is the weighting factor balancing the relative importance of communication and sensing MI, the constraint (22b) ensures the joint eavesdropping MI not to exceed the threshold ${\Gamma _{\rm{E}}}$, constraint (22c) specifies the total power limit, where ${P_{\max }}$ is the power budget. 

\textit{Remark 2.} If the acquired CSI suffers from estimation errors, the actual channel vectors or channel covariance matrices can be modeled as the sum of their estimated values and uncertainty terms. Such CSI uncertainty may affect the beamforming design and the evaluation of the legitimate and eavesdropping performance metrics. For bounded CSI errors, the joint information-leakage constraint can be reformulated in a worst-case manner to ensure security for all possible channel errors within predefined uncertainty sets. For stochastic CSI errors, outage-based robust formulations can be considered as a possible extension.

Problem (22) is non-convex due to the difference-of-concave (DC) structure of the eavesdropping constraint and the rank-1 constraint, which cannot be solved directly. To tackle this problem, we first rewrite $I_{\rm DE}$ in (22b) into 
\begin{align}
	\label{ide1}
\setcounter{equation}{22}
{I_{{\rm{DE}}}} \!=\! {\log _2}\! \left( {1 \!+\! {\rm{Tr}}({\bf{AW}})} \right) \!+\! {\log _2}\! \left(\! {\frac{{{\rm{Tr}}({\bf R}_{\rm ES}{\bf W}) + \sigma _{{\rm{ES}}}^2}}{{(1 \!-\! \alpha ){\rm{Tr}}({\bf R}_{\rm ES}{\bf W}) \!+\! \sigma _{{\rm{ES}}}^2}}} \right),
\end{align}
By substituting $\alpha$ from (\ref{alpha}) into (\ref{ide1}), the joint eavesdropping MI $I_{{\rm{DE}}}$ can be equivalently rewritten as 
\begin{align}
	\label{oi}
{{I_{{\rm{DE}}}} = {f_1}({\bf{W}}) - {f_2}({\bf{W}}}),
\end{align}
where $  f_1({\bf{W}})\!=\!2{\log _2}\! \left( {1 \!+\! {\rm{Tr}}({\bf{AW}})} \right) \!+ \!{\log _2}\! \left( \!{{\rm{Tr}}({\bf R}_{\rm ES}{\bf W}) \!+\! \sigma _{{\rm{ES}}}^2} \right)$\!, ${{f_2}({\bf{W}}) = {{\log }_2}\left( {{\rm{Tr}}({\bf R}_{\rm ES}{\bf W}) + \sigma _{{\rm{ES}}}^2 + \sigma _{{\rm{ES}}}^2{\rm{Tr}}({\bf{AW}})} \right)}$. 

From (\ref{oi}), ${\rm Tr}({\bf A}{\bf W})$ and ${\rm Tr}({\bf R}_{\rm ES}{\bf W})$ are affine functions of ${\bf W}$ and are non-negative over the feasible domain, ensuring that the arguments of the logarithmic functions are positive. Since the logarithmic function is concave and monotonically increasing over the positive domain, both $f_1({\bf W})$ and $f_2({\bf W})$ are concave functions of ${\bf W}$. Therefore, $I_{\rm DE}$ is the difference of two concave functions, thereby exhibiting a DC structure.

To handle this, we adopt the successive convex approximation method. Specifically, $f_1({\bf{W}})$ can be linearized at the current iterate ${\bf{W}}^{(k)}$ using a first-order Taylor expansion. As a result, a convex surrogate function of $f_1({\bf{W}})$ at ${\bf{W}}^{(k)}$ is obtained as 
\begin{align}
\begin{array}{l}
	\!\! f_1({\bf{W}}) \approx \mathord{\buildrel{\lower3pt\hbox{$\scriptscriptstyle\frown$}} 
		\over f_1} ({\bf{W}}) \!=\! f_1({{\bf{W}}^{(k)}}) \!+\! \dfrac{2}{{\ln 2}}\dfrac{{{\rm{Tr}}\left( {{\bf{AW}}} \right) \!-\! {\rm{Tr}}\left( {{\bf{A}}{{\bf{W}}^{(k)}}} \right)}}{{1 + {\rm{Tr}}\left( {{\bf{A}}{{\bf{W}}^{(k)}}} \right)}}\\
	+ \dfrac{1}{{\ln 2}}\dfrac{{{\rm{Tr}}\left( {\bf R}_{\rm ES}{\bf W} \right) - {\rm{Tr}}\left( {{\bf R}_{\rm ES}{{\bf{W}}^{(k)}}} \right)}}{{\sigma _{{\rm{ES}}}^2 + {\rm{Tr}}\left( {{\bf R}_{\rm ES}{{\bf{W}}^{(k)}}} \right)}}.
\end{array}
\end{align}
To handle the rank-1 constraint, we introduce the penalty function ${\left\| {\bf{W}} \right\|_*} - {\left\| {\bf{W}} \right\|_2}$, which encourages the solution to approach a rank-one structure, helps reduce the dependence on additional rank-one recovery procedures, and improves the feasibility of the obtained beamformer. Here, ${\left\|  \cdot  \right\|_*}$ and ${\left\|  \cdot  \right\|_2}$ denote the nuclear norm and spectral norm, respectively. It holds that
\begin{align}
	\label{pen1}
	{\left\| {\bf{W}} \right\|_*} - {\left\| {\bf{W}} \right\|_2} \ge 0,
\end{align}
with equality if and only if ${\bf W}$ is rank 1.

Consequently, the objective (22a) can be reformulated as
\begin{align}
	\label{ob1}
\begin{array}{*{20}{l}}
	{\mathop {\max }\limits_{\bf{W}} \;{I_{\rm{U}}} + \beta {I_{\rm{r}}} - \eta_1 \left( {{{\left\| {\bf{W}} \right\|}_*} - {{\left\| {\bf{W}} \right\|}_2}} \right)},
\end{array}
\end{align}
where $\eta_1$ is the penalty factor.

However, the spectral-norm term in the penalty makes the objective non-convex. By exploiting the first-order Taylor expansion at ${{\bf{W}}^{\left( k \right)}}$, an upper bound of $- {\left\| {\bf{W}} \right\|_2}$ is given by
\begin{align}
	\label{pen2}
	\begin{array}{l}
		\!\!- {\left\| {\bf{W}} \right\|_2} \le {{\mathord{\buildrel{\lower3pt\hbox{$\scriptscriptstyle\frown$}} 
					\over Y} }_1}\left( {\bf{W}} \right)\\
		\!\!=  - {\left\| {{{\bf{W}}^{\left( k \right)}}} \right\|_2}\! -\! {\rm{Tr}}\left[ { \varphi \left( {{{\bf{W}}^{\left( k \right)}}} \right)\!{{\left( {\varphi \left( {{{\bf{W}}^{\left( k \right)}}} \right)} \right)}^{\rm{H}}}\left( {{\bf{W}} - {{\bf{W}}^{\left(k \right)}}} \right)} \right]\!,
	\end{array}
\end{align}
where ${\varphi \left( {\bf{W}} \right)}$ is the eigenvector corresponding to the largest eigenvalue of ${\bf{W}}$. Then, by substituting $ - {\left\| {\bf{W}} \right\|_2}$ with ${{\mathord{\buildrel{\lower3pt\hbox{$\scriptscriptstyle\frown$}} \over Y} }_1}\left( {\bf{W}} \right)$ in (\ref{ob1}), we obtain the following relaxed convex problem
\begin{align}
	\label{opti}
\begin{array}{l}
	\mathop {\max }\limits_{\bf{W}} ~F({\bf W})={I_{\rm{U}}} + \beta {I_{\rm{r}}} - {\eta _1}\left( {{{\left\| {\bf{W}} \right\|}_*} + {{\mathord{\buildrel{\lower3pt\hbox{$\scriptscriptstyle\frown$}} 
					\over Y} }_1}\left( {\bf{W}} \right)} \right)\\
	~{\rm{ s}}{\rm{.t.}}~~ {{{\mathord{\buildrel{\lower3pt\hbox{$\scriptscriptstyle\frown$}} 
					\over f} }_1}({\bf{W}}) - {f_2}({\bf{W}})} \le {\Gamma _{\rm{E}}},\\
	~~~~~~~{\rm{Tr}}\left( {\bf{W}} \right) \le {P_{\max }}, {\bf{W}} \succeq 0,
\end{array}
\end{align}
which is convex and can be efficiently solved using CVX. The algorithm for solving (\ref{opti}) is summarized in \textbf{Algorithm 1}.
\begin{table}[!htbp]
	%\caption{\textbf{Classical table}}%title
	%\centering
	\begin{tabular}{l}% four columns
		\toprule[1pt] %change the first line to \toprule
		{\bf{Algorithm 1:}} {Iterative algorithm for solving problem (\ref{opti})} \\
		\midrule %change the second line to midrule
		{\bf{Input:}} ${\bf H}_{\rm r}$, ${{\bf h}_{\rm U}}$,  ${{\bf{h}}_{\rm{EC}}}$, ${\bf R}_{\rm h_r}$, ${\bf R}_{\rm ES}$, $P_{\rm{max}}$, $\sigma _{\rm{U}}^2$, $\sigma _{\rm{r}}^2$, $\sigma _{\rm EC}^2$, $\sigma _{\rm ES}^2$, \\
		~~~~~~~~~$M,N_{\rm{r}}$, $\beta, \eta_1$, ${\Gamma _{\rm{E}}}$, $\epsilon_1$, $K_{\max}$.\\
		1.~Initialize iteration number $k$ = 0 and ${\bf{w}}^{(0)}$. \\
		2.~\textbf{Repeat:} \\
		3.~~~~~Obtain ${\bf{W}}^{(k+1)}$ by solving (\ref{opti}) with given ${\bf{W}}^{(k)}$. \\
		4.~~~~~Construct the beamformer ${\bf{w}}^{(k+1)}$ by eigenvalue decomposition. \\
		5.~~~~~Compute the penalized objective value $F^{(k+1)}$. \\
		6.~~~~~Let $k=k+1$.  \\
		7.~\textbf{Until} $\dfrac{|F^{(k)}-F^{(k-1)}|}{\max\{1,|F^{(k-1)}|\}}\le \epsilon_1$ or $k\ge K_{\max}$. \\
		{\bf{Output:}} $\bf{w}$. \\
		\bottomrule[1pt] %change the third line to bottomrule
	\end{tabular}
\end{table}
\subsection{Convergence and Complexity Analysis}
\subsubsection{Convergence Analysis}
In Algorithm 1, the convex surrogate problem (\ref{opti}) is solved at each iteration, and the solution obtained in the previous iteration remains feasible for the current surrogate problem due to the tightness of the first-order approximation. Therefore, the penalized objective value is non-decreasing over iterations. Since the feasible set is compact due to the constraint in (\ref{opti}), the penalized objective value is upper-bounded. Consequently, the generated objective sequence is guaranteed to converge \cite{conver}.

\subsubsection{Complexity Analysis}
We analyze the worst-case computational complexity of the proposed algorithm via the interior-point method \cite{collude}. The primary computational burden comes from solving problem (\ref{opti}), which contains $\phi=M^2$ optimization variables, a constant number of affine constraints, and one LMI constraint of size $M$. Therefore, the number of iterations required is $O( {\sqrt M \ln \left( {1/\epsilon_2 } \right)} )$, where $\epsilon_2$ represents the solution accuracy. The computational cost per interior-point iteration is given by $O\left( {\phi \left( {{M^3} + \phi {M^2} + {\phi ^2}} \right)} \right)$. Finally, the total computational complexity becomes $O\left( {{M^{6.5}}\ln \left( {1/\epsilon_2 } \right)} \right)$.
%------------------------------------------------------------------------------------
\section{Numerical Results} \label{4jie}      
In this section, we provide numerical analysis to evaluate the effectiveness of the proposed secure ISAC design. The BS is equipped with $M=6$ transmit antennas and $N_{\rm r}=2$ receiving antennas, with the power budget $P_{\rm max}=25$ dBm. We set all noise power as 0.01 W. The joint eavesdropping MI threshold is established at ${\Gamma _{\rm{E}}} = 2$ bps/Hz. Unless otherwise specified, the convergence tolerance and maximum number of iterations are set to $\epsilon_1=10^{-4}$ and $K_{\rm max}=10$, respectively.

\begin{figure}[t]
	\centering
	\includegraphics[width=3.0in]{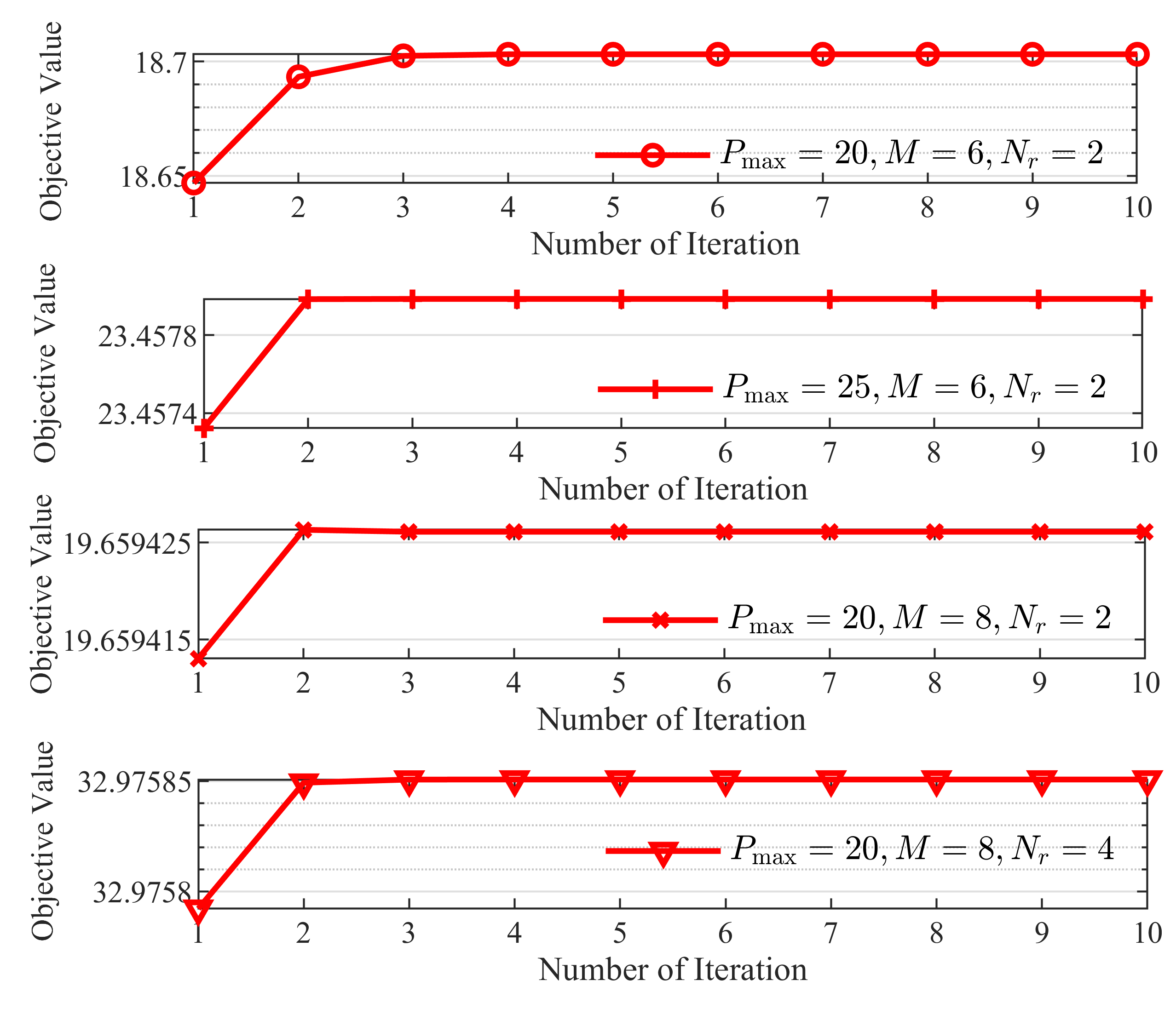}
	\caption{Convergence behavior of the proposed algorithm.}
	\label{Convergence}
\end{figure}

Fig. \ref{Convergence} shows the convergence performance of the proposed algorithm under different parameter settings. It can be observed that the objective value converges rapidly and remains constant for all settings, indicating fast convergence. Moreover, the converged objective value increases with the power budget and the numbers of transmit and receiving antennas, since the increased degrees of freedom enhance the legitimate performance under the same security constraints.
\begin{figure}[t]
	\centering
	\includegraphics[width=3.0in]{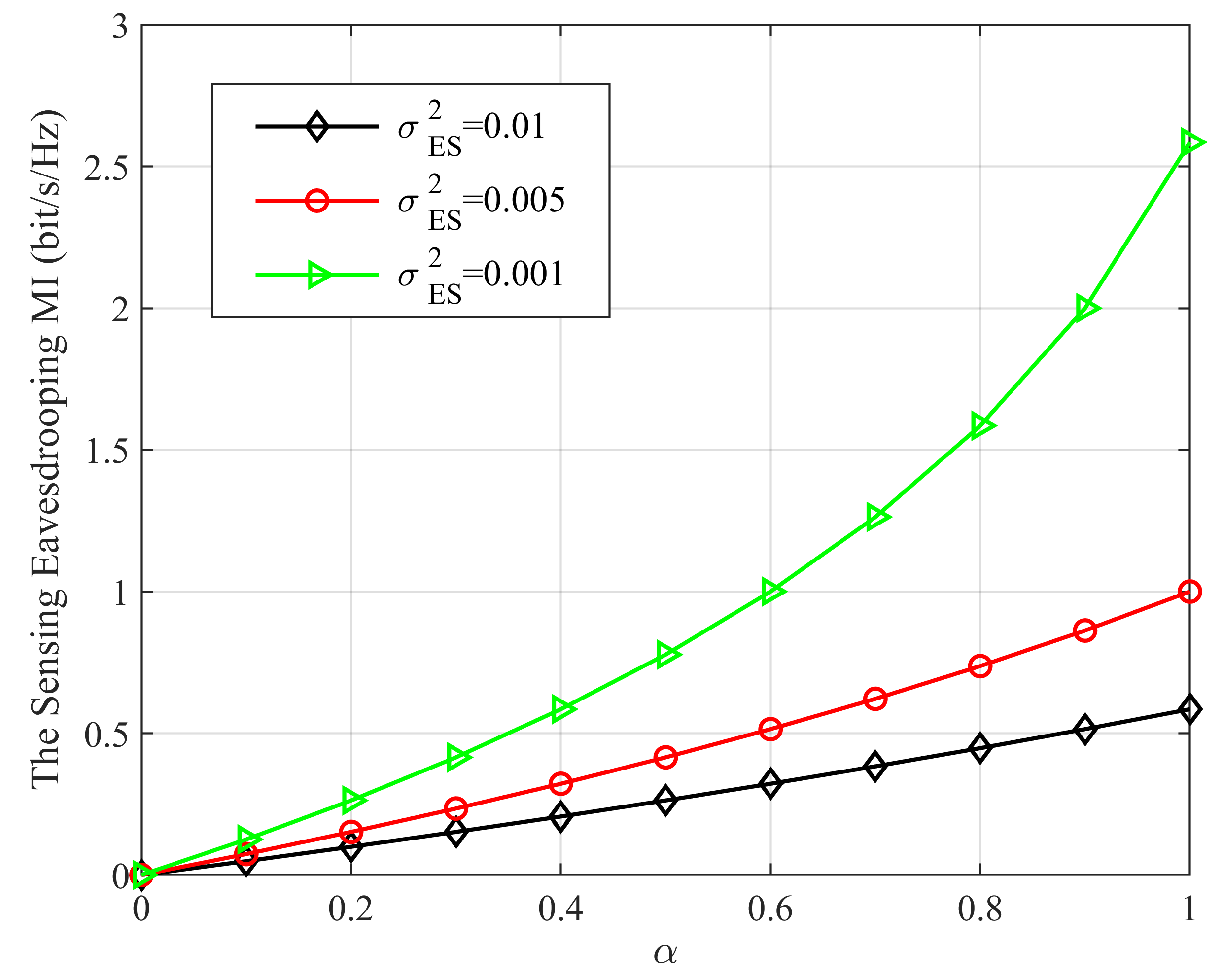}
	\caption{The sensing eavesdropping MI $I_{\rm ES}$ with respect to $\alpha$ ($M=8$, $P_{\rm{max}}=30~{\rm{dBm}}$, ${\bf R}_{\rm{ES}}=0.01{\bf I}$).}
	\label{AlphaMono}
\end{figure}

Fig. \ref{AlphaMono} shows the sensing eavesdropping MI $I_{\rm ES}$ with respect to the RSR $\alpha$, under a fixed transmit beamformer. It can be observed that $I_{\rm ES}$ increases monotonically with $\alpha$, which numerically validates \textit{Lemma 1}, thereby demonstrating the impact of communication eavesdropping on sensing information leakage. Moreover, a higher sensing eavesdropping SINR results in more severe sensing information leakage. 

Fig. 4 plots the legitimate MI and the joint eavesdropping MI with respect to the eavesdropping MI threshold $\Gamma_{\rm E}$ under different communication eavesdropping recovery conditions. Specifically, the proposed secure beamforming design, where the symbol recovery at the CEve is characterized by the RSR $\alpha$, is compared with three benchmarks. The \textbf{SEve-Only} scheme follows the sensing-security setting in \cite{Zou}, where only the SEve is considered and perfect signal knowledge is assumed at the SEve. The \textbf{Perfect-Recovery} scheme corresponds to $\alpha=1$ in \textit{Lemma 2}, extending the perfect signal-knowledge assumption in \cite{Zou} to the cooperative dual-eavesdropper scenario. The \textbf{No-Recovery} scheme corresponds to $\alpha = 0$, where no symbol component is recoverable at the CEve, and is adapted from the non-cooperative dual-eavesdropper design in \cite{ren1}.
\begin{figure}[t]
	\centering
	\subfigure[RSR and joint eavesdropping MI $I_{\rm DE}$]{
		\includegraphics[width=0.9\linewidth]{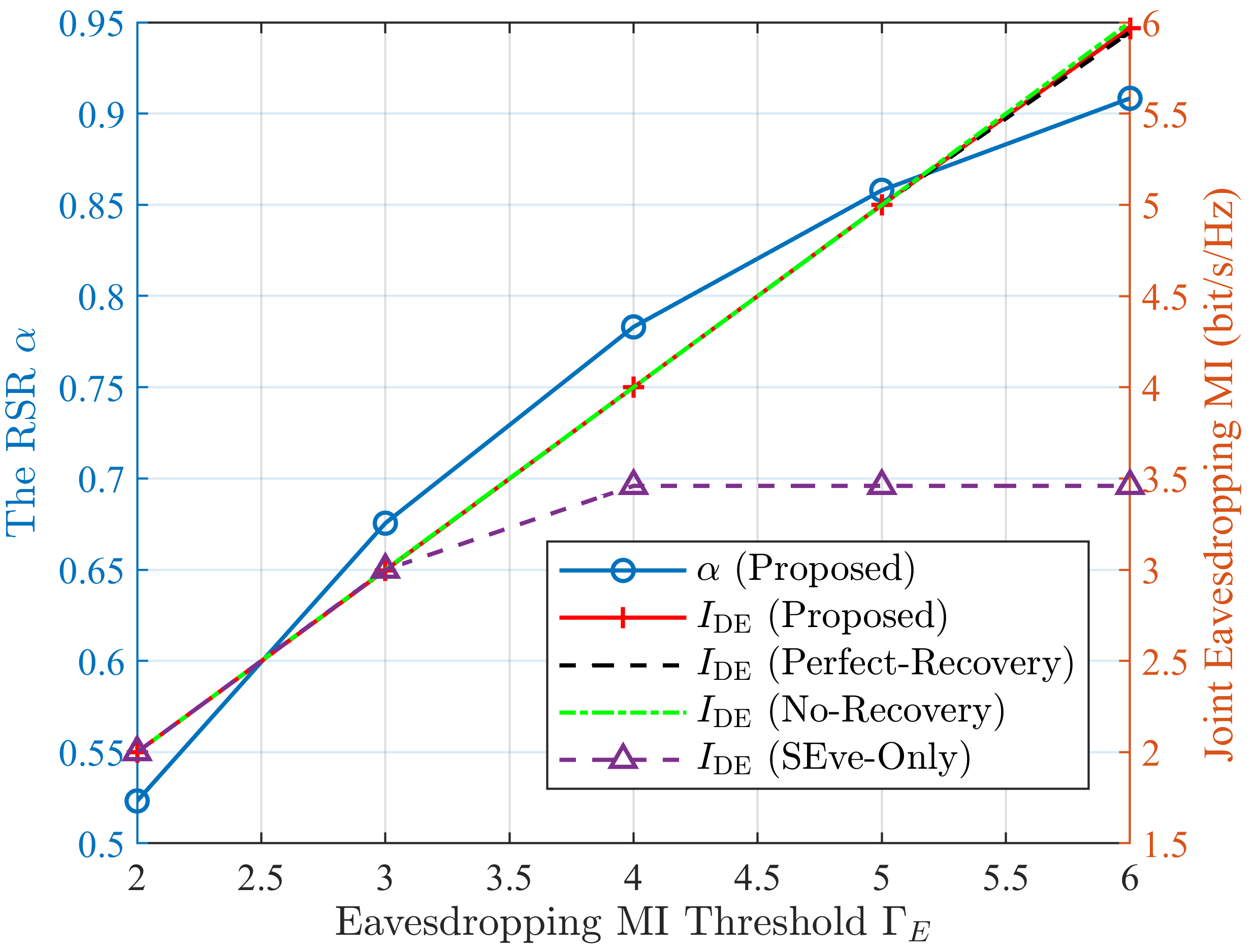}
		\label{GammaE_a}
	}
	\vspace{3pt}	
	\subfigure[Weighted legitimate MI]{
		\includegraphics[width=0.88\linewidth]{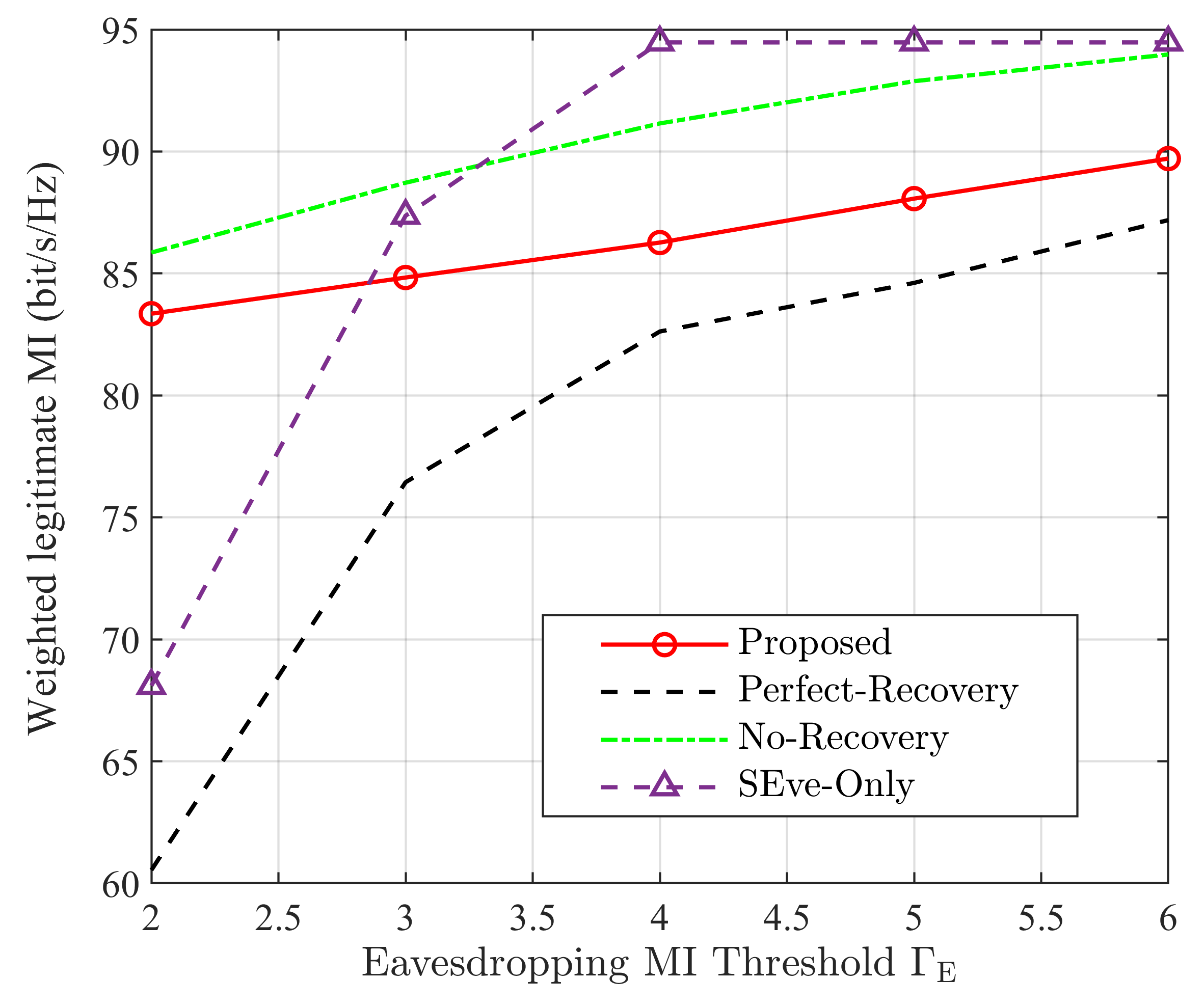}
		\label{GammaE_b}
	}
	\caption{Joint eavesdropping and legitimate performance with respect to the eavesdropping MI threshold $\Gamma_{\rm E}$ ($M=6$, $N_{\rm r}=2$, ${\bf R}_{\rm h_r}=0.01{\bf I}$).}
	\label{ww1}
\end{figure}

From Fig. \ref{GammaE_a}, it is observed that the joint eavesdropping MI of the proposed scheme, the Perfect-Recovery scheme, and the No-Recovery scheme closely approaches the predefined threshold $\Gamma_{\rm E}$. Moreover, as $\Gamma_{\rm E}$ increases, the optimized $\alpha$ monotonically increases, indicating that a less restrictive eavesdropping constraint allows a larger recoverable signal component at the CEve, thereby increasing sensing information leakage. For the SEve-Only baseline, $I_{\rm DE}$ refers to the sensing eavesdropping MI, since no CEve is considered. It first increases and then remains nearly unchanged because the sensing-leakage constraint becomes inactive when $\Gamma_{\rm E}$ is sufficiently large.

Fig. \ref{GammaE_b} shows that the legitimate MI of the proposed scheme, the Perfect-Recovery scheme, and the No-Recovery scheme monotonically increases with $\Gamma_{\rm E}$. For a given $\Gamma_{\rm E}$, the No-Recovery scheme achieves the highest legitimate MI, followed by the proposed scheme, while the Perfect-Recovery scheme yields the lowest legitimate MI. This is because a smaller $\alpha$ corresponds to weaker sensing eavesdropping capability, which relaxes the information-leakage constraint, enlarging the feasible design region and thereby enhancing the legitimate communication and sensing performance. For the SEve-Only baseline, the legitimate MI also remains stable for large $\Gamma_{\rm E}$, since the sensing-leakage constraint becomes inactive. These results validate the effectiveness of the proposed scheme in characterizing and exploiting the coupled communication and sensing information leakage.
\section{Conclusion}\label{6jie}                               
This paper investigated the joint communication and sensing security in low-altitude ISAC systems with cooperative CEve and SEve. To this end, we introduced the RSR to quantify the coupling between the communication eavesdropping recovery and the resulting sensing information leakage at the SEve. We also developed a joint eavesdropping MI metric to characterize this coupling and designed the transmit beamformer that maximizes the overall secure performance of legitimate communication and sensing while limiting joint information leakage. Finally, simulation results showed that the proposed scheme converges rapidly and that the RSR critically influences both sensing information leakage and overall security performance.

\newpage

\vfill

\end{document}